\documentclass[sigplan,screen]{acmart}
\AtBeginDocument{%
  \providecommand\BibTeX{{%
    \normalfont B\kern-0.5em{\scshape i\kern-0.25em b}\kern-0.8em\TeX}}}
\setcopyright{acmlicensed}
\copyrightyear{2024}
\acmYear{2024}
\acmDOI{XXXXXXX.XXXXXXX}

\acmConference[SETN 2024]{September 11-13, 2024}{Piraeus, Greece}

\acmISBN{978-1-4503-XXXX-X/18/06}

\usepackage{fancyhdr}
\usepackage{graphicx}
\usepackage[ruled,vlined]{algorithm2e}

\begin{document}

\title{Multi-robot maze exploration using an efficient cost-utility method}

\author{Manousos Linardakis}
\affiliation{%
  \textit{\institution{Department of Informatics and Telematics}}
  \textit{\institution{Harokopio University of Athens}}
  \city{Athens}
  \country{Greece}
}
\email{manouslinard@gmail.com}

\author{Iraklis Varlamis}
\affiliation{%
  \textit{\institution{Department of Informatics and Telematics}}
  \textit{\institution{Harokopio University of Athens}}
  \city{Athens}
  \country{Greece}
}
\email{varlamis@hua.gr}

\author{Georgios Th. Papadopoulos}
\affiliation{%
  \textit{\institution{Department of Informatics and Telematics}}
  \textit{\institution{Harokopio University of Athens}}
  \city{Athens}
  \country{Greece}
}
\email{g.th.papadopoulos@hua.gr}

\begin{abstract}
\textbf{In the field of modern robotics, robots are proving to be useful in tackling high-risk situations, such as navigating hazardous environments like burning buildings, earthquake-stricken areas, or patrolling crime-ridden streets, as well as exploring uncharted caves. These scenarios share similarities with maze exploration problems in terms of complexity. While several methods have been proposed for single-agent systems, ranging from potential fields to flood-fill methods, recent research endeavors have focused on creating methods tailored for multiple agents to enhance the quality and efficiency of maze coverage. The contribution of this paper is the implementation of established maze exploration methods and their comparison with a new cost-utility algorithm designed for multiple agents, which combines the existing methodologies to optimize exploration outcomes. Through a comprehensive and comparative analysis, this paper evaluates the performance of the new approach against the implemented baseline methods from the literature, highlighting its efficacy and potential advantages in various scenarios. The code and experimental results supporting this study are available in the following \href{https://github.com/manouslinard/multiagent-exploration/}{repository}.
}
\end{abstract}



\received{13 April 2024}
\received[accepted]{15 July 2024}

\maketitle

\section{Introduction}
Maze exploration has long interested researchers, serving as a valuable tool in understanding the cognitive abilities of various organisms, like mice 
and more recently in assessing the artificial intelligence capabilities of robotic systems \cite{robotics_maze2, robotics_maze1}. While algorithms tailored for single-agent maze exploration are widely employed \cite{singlerobot_path_img, singlerobot_path}, adapting them for use in multi-agent systems \cite{multiagent_ai} poses ongoing challenges, driving the need for innovative solutions.

In scenarios where the maze layout is predetermined \cite{astar_knownmaze}
agents typically rely on established path-planning algorithms to navigate towards predefined objectives, such as exits. However, when navigating in
unknown maze configurations \cite{hedac_maze, cu_jgr}, agents must employ sensor-based exploration strategies to gather information about their surroundings before devising navigation plans. This necessitates addressing a multitude of challenges related to sensing, exploration, and path-finding.

Mapping unknown mazes for exploration demands strategic approaches. Robots navigate through the maze, recording their journey and observations to construct a map as they progress. Utilizing sensors, they detect walls and obstacles, making informed decisions about their next move.
Once a sufficient portion of the maze is explored and mapped, robots strategize their routes toward their objectives, whether it be reaching the exit or locating specific targets within the maze. This involves analyzing the mapped environment and selecting the optimal path forward.

Using multiple agents to explore the maze accelerates the process and enhances the quality of the map, but it also introduces challenges such as agent communication for information sharing and preventing collisions or path obstructions.

Our proposed cost-utility approach has demonstrated efficiency in addressing these challenges, facilitating a more effective full-maze exploration in terms of total communication between agents, total distance traveled, and total time needed for the exploration. 
The main contributions of the proposed work can be summarized in the following:
\begin{itemize}
    \item A detailed review of the recent exploration approaches using multiple agents.
    \item A new efficient algorithm for maze exploration by multiple agents, which combines the nearest frontier method and a new cost-utility function.
    \item An extended implementation of various multi-agent maze exploration methods.
    \item A comparative evaluation of the algorithms, using various metrics that examine different aspects of the exploration task (e.g. time, distance, computational cost).
\end{itemize}

Section \ref{sec:relwork} that follows provides an overview of the related work in the field of maze and area exploration with emphasis on multi-agent approaches. Section \ref{sec:proposed_approach} details the proposed approach and Section \ref{sec:experimental_eval} explains the experimental evaluation process that we applied. Section \ref{sec:result_plots} discusses the achieved results and finally Section \ref{sec:conclusions} presents the main conclusions of this work and the next steps.

\section{Related Work}
\label{sec:relwork}
Research into multiple-agent exploration has yielded diverse methodologies and algorithms. A simple yet effective approach is the nearest-frontier method introduced by Yamauchi \cite{yamauchi1998frontier}. This approach selects the shortest path to the nearest frontier. A frontier is defined as the border between explored and unexplored areas during coverage.
Despite the efficiency of nearest-frontier, more advanced exploration methods have emerged. Gul et al. \cite{gul2021novel} and RSIS International \cite{rsis_international_2016} categorize multi-robot exploration into various approaches (also see Table \ref{tab:relwork}):

\begin{itemize}
\item \textbf{Socratic Approach}: Involves a cyclic process of sensing, positioning, decision making, and movement in which agents 
execute actions while maintaining a global map for efficient coordination.
\item \textbf{Cellular/Stochastic Approach}: Draws inspiration from natural systems like ant colonies and the human immune system to enhance exploration.
\item \textbf{Market-Based Approach}: Applies socio-economic principles, such as the "Social potential field", to optimize resource utilization and task allocation among multiple agents.
\item \textbf{Tree and Graph Approach}: Reshapes maze exploration as a tree/graph problem.
\item \textbf{Hybrid Approach}: Combines various techniques and algorithms for efficient maze exploration.
\item \textbf{Deterministic Approach}: Utilizes the coordination of multiple robots to explore the area by a centralized technique, exemplified by frontier-based exploration.
\end{itemize}



The research by RSIS \cite{rsis_international_2016} also describes approaches for maze pathfinding. Pathfinding differs from maze exploration since the goal is to find a solution to the maze, rather than to explore it completely. Some approaches for maze path finding include the multi-hop approach discussed by Jung et al. \cite{multihop_pathfind} and a variation of the tree and graph approach flood-fill by Tjiharjadi et al. \cite{tjiharjadi2022design}. The latter underscores the prevalence of single-agent maze exploration/pathfinding and investigates the utility of multi-agent pathfinding in maze environments. Another systematic survey by Tjiharjadi et al. \cite{tjiharjadi2022systematic}, that once again highlights the use of flood fill and other pathfinding algorithms ($A^*$ and Dijkstra) for maze solving, proves that multi-agent pathfinding is predominantly applied in static environments with known target locations. 

Other exploration methods, as discussed by Sharma and Tiwari \cite{sharma2016survey}, encompass a wide range of techniques. These include cost-utility approaches, flood fill algorithms, genetic algorithms \cite{gen_algorithms}, graph exploration techniques, and Rapidly Exploring Random Trees (RRT) \cite{rrt2}.


\subsection{Socratic Approaches}
\label{subsec:socratic_approach}
\begin{figure}[h]
    \centering
    \includegraphics[width=0.45\textwidth]{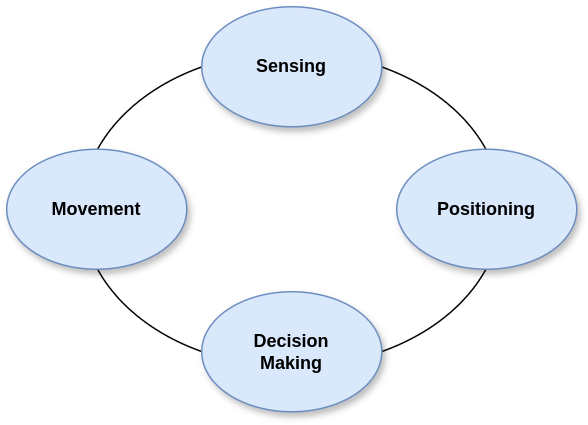}
    \caption{Abstract Schema of Socratic Approach.}
    \label{fig:socratic}
    \Description{Abstract Schema of Socratic Approach.}
\end{figure}

The Socratic approach \cite{reid_socratic} (also known as heuristic) in mapping tasks, uses an online method for likelihood maximization employing hill climbing techniques. This method aims to enhance overall utility by minimizing the overlap in information gathered by robots. In mapping and exploration tasks, computations are predominantly distributed, with each robot processing its own sensor data to create local maps which are then integrated globally by a central merging mechanism to generate a reliable map. Each robot measures three key elements from sensor data: a Maximum Likelihood Estimate (MLE) for its location, an MLE for obstacles in its vicinity, and a posterior density indicating the true position. Whenever a map update is received from the central mapper, robots construct new bids and send them to the central executive, which assigns tasks based on the received bids. An example of the socratic approach can be found in Figure \ref{fig:socratic}.

\subsection{Cellular Approaches}
\label{subsec:cellular_approach}
Wagner et al. \cite{antwalk} introduce an approach inspired by ants, leveraging their method of food communication through pheromones or chemical signals to inform others. Similarly, robots are equipped with a mechanism for depositing signals (pheromones), during coverage and exploration tasks, which gradually decay over time.

Butler et al. \cite{seed_sowl} propose a cellular coverage algorithm for multiple agents using seed-sowing. The seed-sowing technique involves traversing each cell systematically, with robots covering the area in stripes. During the exploration process, robots maintain individual internal maps, and process handlers monitor incoming instructions. Robots are positioned either in parallel to each other or start their exploration in a perpendicular arrangement.

\subsection{Market-Based Approaches}
\label{subsec:market_approach}
HEDAC \cite{hedac_maze, hedac_og} is a robust algorithm for both area and maze exploration. By leveraging potential fields to compute attractive forces, HEDAC excels in prioritizing the exploration of unknown regions while mitigating collisions. Specifically, HEDAC creates an attractive force for unexplored areas and a less attractive force for obstacles and other agents. Moreover, HEDAC has demonstrated efficacy in uncertain conditions, such as variations in sensor readings and unpredictable environmental fluctuations, as documented in \cite{ivic2020motion}.

A similar approach to HEDAC that uses Voronoi diagrams to split the maze into subareas is presented in \cite{zheng2022distributed}. Each subarea is then explored by an agent using a temperature field-induced control strategy.
Similarly, in \cite{olcay2020sensor} an APF method is designed to navigate around obstacles of varying shapes while autonomously mapping the entire area, by identifying inaccessible domains. The agent's interaction is limited only with agents within its communication bandwidth.

\begin{figure}[h]
    \centering
    \includegraphics[width=0.47\textwidth]{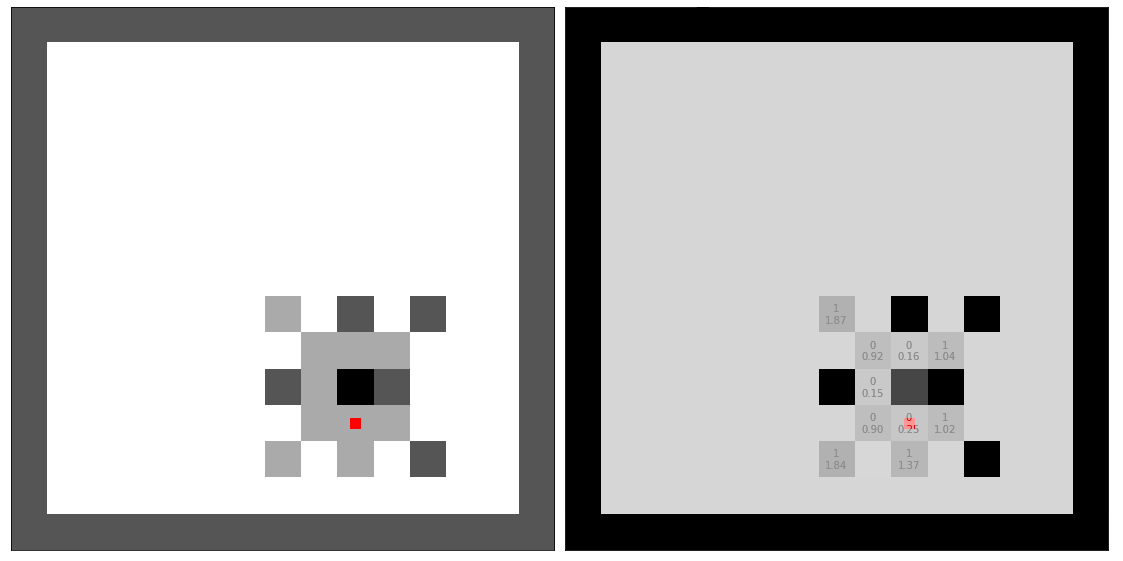}
    \caption{Example of potential fields usage in maze exploration.}
    \Description{Example of potential fields usage in maze exploration.}
    \label{fig:market_based}
\end{figure}

An example of potential fields usage can be found in Figure \ref{fig:market_based}. The neighboring non-obstacle (light gray) cell with the highest attractive force becomes the next target (indicated by a red dot) for the agent (black cell on the left image).

\subsection{Tree \& Graph Approaches}
\label{subsec:tree_graph_approach}
Cabrera et al. \cite{flood_fill_explore} introduce a flood-fill algorithm tailored for maze exploration. The algorithm is designed to decrease exploration time and minimize robot traversal distance by coordinating robot movements. This decentralized coordination relies on robots deploying active landmarks at explored vertices to guide their exploration. 


The method proposed in \cite{multiagent_graph}
for mapping undirected graphs using multiple synchronous operators involves a colony of agents departing from a common starting vertex and dispersing to explore the graph (maze). This approach treats a centralized tree for the graph as a closed loop, which is verified as agents analyze common vertices. By incorporating an active exploration process, where agents dynamically seek rendezvous tasks among each other, robust mapping is achieved to validate the graph hypothesis. 

\begin{figure}[h]
    \centering
    \includegraphics[width=0.47\textwidth]{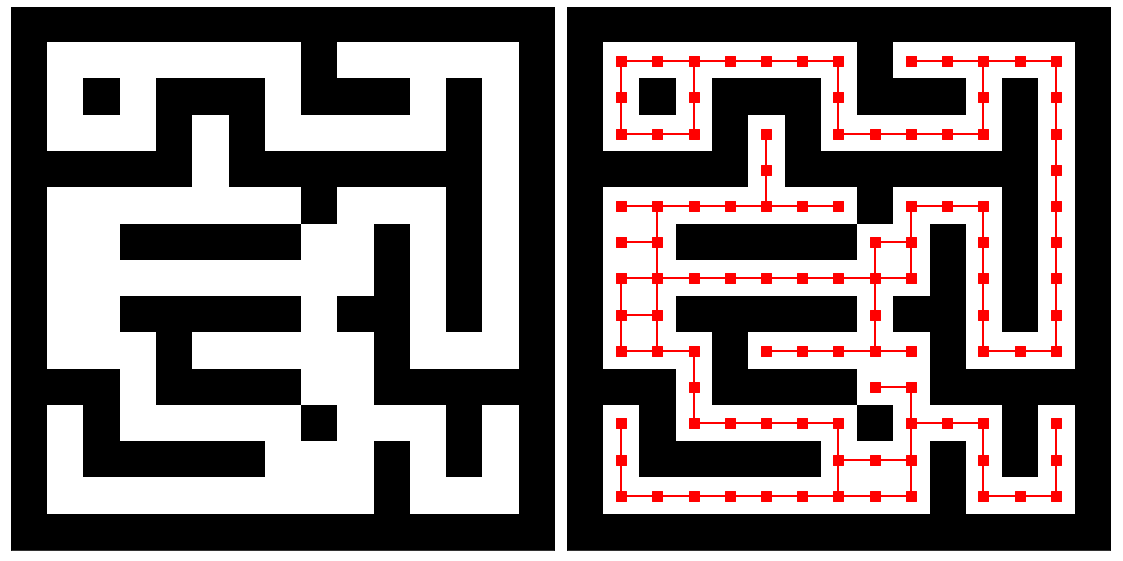}
    \caption{Maze converted to Graph (Tree \& Graph Approach).}
    \Description{Maze converted to Graph (Tree \& Graph Approach).}
    \label{fig:tree_graph}
\end{figure}

An example of converting a maze to a graph is illustrated in Figure \ref{fig:tree_graph}. The left side of the figure shows the original maze, where black blocks represent obstacles. The right side demonstrates the graph representation of the maze, where the red line and square markers indicate the nodes and edges of the graph, representing the available paths.

\subsection{Hybrid Approaches}
\label{subsec:hybrid_approach}
Gul et al. \cite{gul2021multi} present a Hybrid Stochastic Optimizer (HSO) for multi-robot space exploration, combining Coordinated Multi-robot Exploration (CME) with Arithmetic Optimization (AO) techniques to optimize utility. Extensive testing across diverse map complexities demonstrates notable enhancements in exploration metrics, including expanded explored areas and reduced search times. While excelling in highly complex environments, the algorithm's performance in low-complexity settings remains comparable to other methods, suggesting potential for further refinement.
In another study \cite{gul2021novel}, the authors propose a novel framework integrating CME with cellular Whale Optimization Algorithm (WOA) techniques, mimicking whale behavior. This hybrid approach dynamically adjusts frequency parameters to optimize search space exploration, showing superior efficacy compared to contemporary methods, including CME-WOA, CME-GWO, and CME-SineCosine, across varied environmental complexities.

\subsection{Determinstic Approaches}
\label{subsec:deterministic_approach}
Cost-utility methods are a focal point in multi-agent maze exploration research. These methods typically enhance the deterministic nearest-frontier approach by incorporating an additional utility function. In essence, the cost component represents the distance to the frontier, while the utility function offers a customized criterion for selecting the most favorable frontier. Researchers strive to minimize the cost function while simultaneously maximizing the utility function to optimize exploration efficiency.

The work by Marjovi et al. \cite{cu_mnm_paper} introduces a cost-utility approach that efficiently explores mazes while detecting fires. The utility function is based on the distance of the frontier from all robots. Decentralized control facilitates communication and collaborative decision-making among robots.
Another cost-utility approach, proposed in \cite{cu_jgr}, employs as its utility function the expected information gain in goal cells for efficient exploration.

A  computationally efficient frontier allocation method that promotes a well-balanced spatial distribution of robots within the environment is presented in \cite{cu_bso}. It employs wavefront propagation \cite{wavefront} ascending from each frontier to quickly generate a distance matrix (cost-matrix) for all cells and subsequently selects the frontiers with the fewer nearby robots. The use of wavefront propagation underscores reduced communication requirements among robots. Burgard et al. \cite{burgard2005coordinated} propose an efficient cost-utility approach that yields favorable results in practical experiments, showcasing its effectiveness in real-world settings.


\begin{table}[!htb]
\caption{The main approaches for multi-agent maze exploration. }
\label{tab:relwork}
\resizebox{\columnwidth}{!}{%
\begin{tabular}{|c|c|c|c|l}
\cline{1-4}
\textbf{Approach} & \textbf{Methods} & \textbf{Features} & \textbf{Citations} &  \\ \cline{1-4}
Socratic & MLE & Central Mapper & \cite{reid_socratic} &  \\ \cline{1-4}
Cellular & \begin{tabular}[c]{@{}c@{}}Ant-Walk, \\ Seed-Sowing\end{tabular} & \begin{tabular}[c]{@{}c@{}}Nature based\\ Algorithms\end{tabular} & \cite{seed_sowl}, \cite{antwalk} &  \\ \cline{1-4}
Market-Based & \begin{tabular}[c]{@{}c@{}}HEDAC, \\ APF\end{tabular} & \begin{tabular}[c]{@{}c@{}}Potential Field, \\ Temperature\end{tabular} & \begin{tabular}[c]{@{}c@{}}\cite{hedac_maze}, \cite{hedac_og},\\ \cite{olcay2020sensor}, \cite{renzaglia2010potential}, \\ \cite{zheng2022distributed} \end{tabular} 
 &  \\ \cline{1-4}
Tree \& Graph & Flood Fill & \begin{tabular}[c]{@{}c@{}}Local or Global\\ Tree is built\end{tabular} & \cite{flood_fill_explore}, \cite{multiagent_graph} &  \\ \cline{1-4}
Hybrid & \begin{tabular}[c]{@{}c@{}}FMH-WOA,\\ CME-AOA\end{tabular} & \begin{tabular}[c]{@{}c@{}}Combination of\\ Algorithms\end{tabular} & \cite{gul2021multi}, \cite{gul2021novel} &  \\ \cline{1-4}
Deterministic & \begin{tabular}[c]{@{}c@{}}Cost-utility \\ \& NF\end{tabular} & \begin{tabular}[c]{@{}c@{}} Frontier \\ Exploration \end{tabular} & \begin{tabular}[c]{@{}c@{}}\cite{cu_bso}, \cite{burgard2005coordinated} \\ \cite{cu_jgr}, \cite{cu_mnm_paper} \end{tabular} &  \\ \cline{1-4}
\end{tabular}
}
\end{table}

The proposed method applies to robotic agents that aim to fully explore and map a previously unknown static maze. It is a deterministic approach aiming at optimizing frontier selection and enhancing overall maze exploration efficiency with multiple agents. The method introduces a novel cost-utility function, building upon the nearest-frontier concept, to help each agent choose the best target at each step and explore the maze efficiently. The results show that it outperforms potential field algorithms like HEDAC and other deterministic approaches, which are the current state-of-the-art, by adding a small computational overhead for the computation of the new cost-utility function.

\section{Proposed Approach}
\label{sec:proposed_approach}
Before going into the details of the proposed approach and algorithm it is necessary to define the main entity of our method, \textit{the agent}.

\subsection{Definition of the robotic agent}
\label{subsubsec:agent_class}

In our approach, we assume an agent class that embodies the behavior of individual agents within the maze environment. The agent can move up, down, left, and right. To prevent collisions, agents are restricted from moving into occupied cells, whether by other agents or obstacles.

Each agent maintains a personal map representing the explored areas of the maze. This map is continually updated based on information gathered from other agents in each round. By effectively sharing information about the explored maze between agents, we replicate typical multi-agent behavior/communication for maze exploration. Specifically, agents share their personal maps, including recently explored cells, with each other. This collaborative approach allows them to make informed decisions about which frontier is best to explore based on the latest information.

Additionally, the agent's map is updated as it explores new areas within its sensor range. By default, the sensor range extends to two blocks in all orthogonal and diagonal directions, mimicking the sensing capabilities of real-world robots, which are typically equipped with sensors on all sides. 

\begin{figure}[h]
    \centering
    \includegraphics[width=0.4\textwidth]{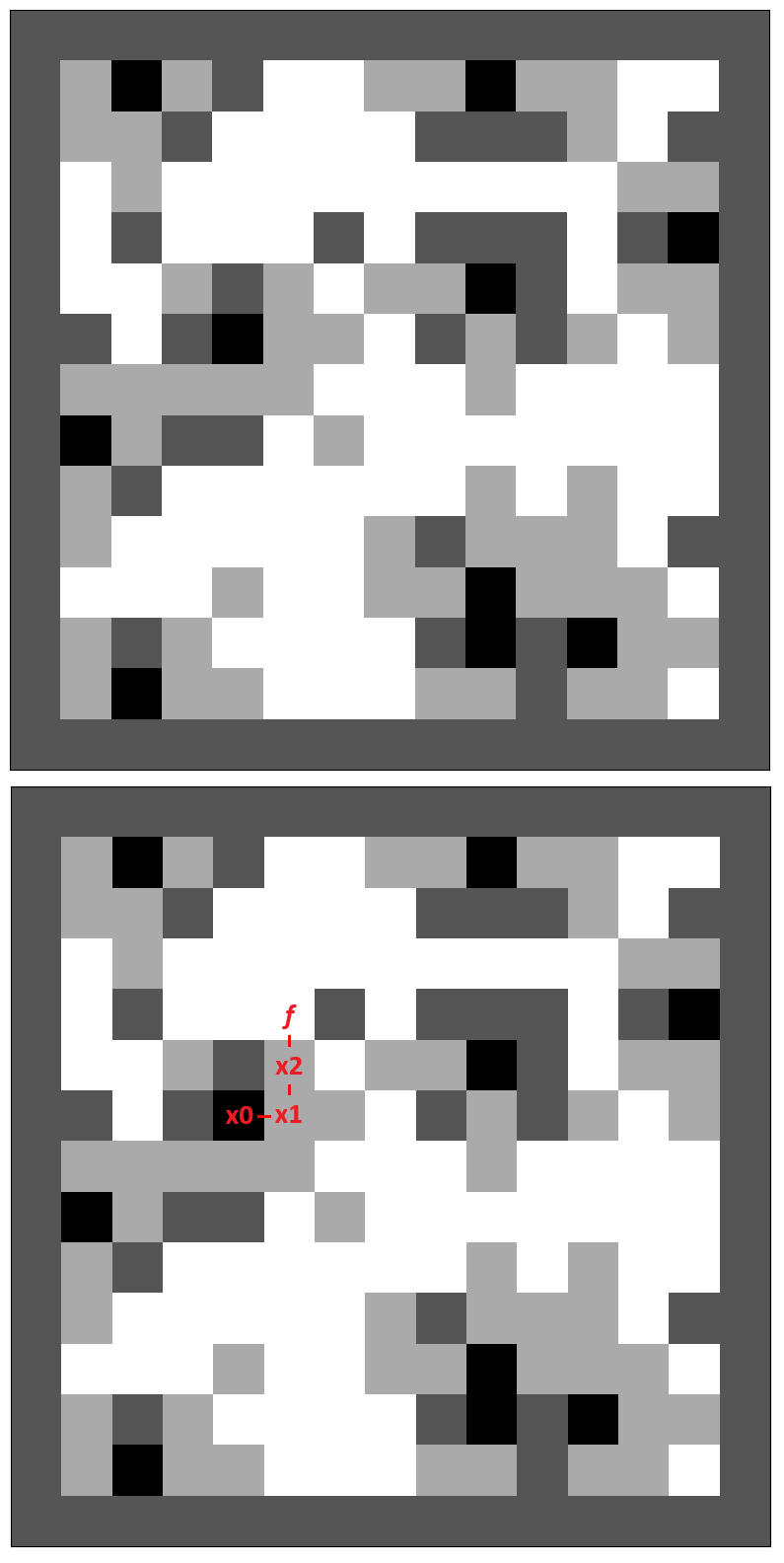}
    \caption{Visualization of the exploration process. 
    The black cells represent the agents. The light gray cells denote free space, and the darker gray cells represent obstacles. }
    \label{fig:agent_view}
    \Description{Visualization of the proposed method's exploration process.}
\end{figure}

Figure \ref{fig:agent_view} (top image) presents the explored map of an agent at the beginning, following the initial information exchange between agents. The map is displayed with view ranges of two blocks. Notably, if an obstacle is detected, the view is blocked beyond it, to simulate real-sensor behavior. 
The agent class also stores essential parameters required for exploration, including the agent's goal (denoted with a red $f$ in bottom image of Figure \ref{fig:agent_view}), the next position based on the shortest wavefront path ($x_1, x_2$) to the target, the current agent position ($x_0$), and the view range. These parameters enable us to implement a realistic agent behavior and implementation scenarios.

Initially, the algorithm employs wavefront propagation \cite{wavefront}, ascending from the agent position to
efficiently compute the lengths of the shortest paths to reach the frontiers. The nearest frontier is selected among the unexplored cells and the selection is based on distance. If multiple frontiers share the same minimum distance from the agent, for each frontier $a_i$ in the nearest frontier list, we compute the utility function. The frontier with the highest utility score is selected as the next target cell for the agent. 

The proposed approach extends 
the methodology outlined in \cite{cu_mnm_paper}, which prioritizes nearest-frontier computation, by introducing an improved cost-utility function for the resolution for equidistant nearest-frontiers.
Our method also ensures that all the agents have different goals if the number of unexplored frontiers is greater than or equal to the number of agents.
Finally, it includes mechanisms to avoid collisions and path blocking in the case of multiple agents, which are explained in the following.

\subsection{Goal selection}
\label{subsec:goal_selection}
The computational requirements of the proposed approach primarily involve the calculation of wavefront and the utility function for selecting the optimal frontier. While using a utility introduces slightly higher initial computational demands, it leads to overall faster exploration by avoiding inefficient frontier choices. However, this trade-off is minimal compared to other methods, due to the fast execution of wavefront propagation, as can be seen in experiment results (Section \ref{sec:result_plots}).

The utility function of the proposed method combines elements from the cost-utility functions presented in \cite{cu_mnm_paper} (later referenced as $u_{mnm}$) and the approach presented in \cite{cu_jgr}. Instead of merely considering the neighboring cells (in view range) of the goal cell in the cost-utility function, as \cite{cu_jgr} does, our method considers the neighboring cells along the whole wavefront path from the current position of the agent to its goal cell (later referenced as $u_{jgr}$). This comprehensive calculation provides a more accurate estimation of the expected explored cells and improves the overall efficiency of the maze exploration task as demonstrated in the experimental evaluation. 
Specifically, the formula for this new cost-utility function is as follows:

\begin{equation}
\text{utility}(f) = N(u_{mnm}(f)) + \lambda \cdot N(u_{jgr}(f))
\label{eq:utility}
\end{equation}

\noindent where $f$ represents the nearest frontier cell under examination, and $N(x)$ denotes the min-max normalization function, ensuring values fall within the range [0, 1]. 
If cell $x_0$ is the current position of agent $x$, the path $P$ is the sequence $x_1, ..., x_i, ..., x_n$ of cells that the agent has to cross in order to reach the target $f$. These cells are explored and accessible but within the view range of the agent, there may be several unexplored cells that will be explored if the agent selects the specific frontier $f$ and the respective path $P$.

\begin{figure}[h]
    \centering
    \includegraphics[width=0.47\textwidth]{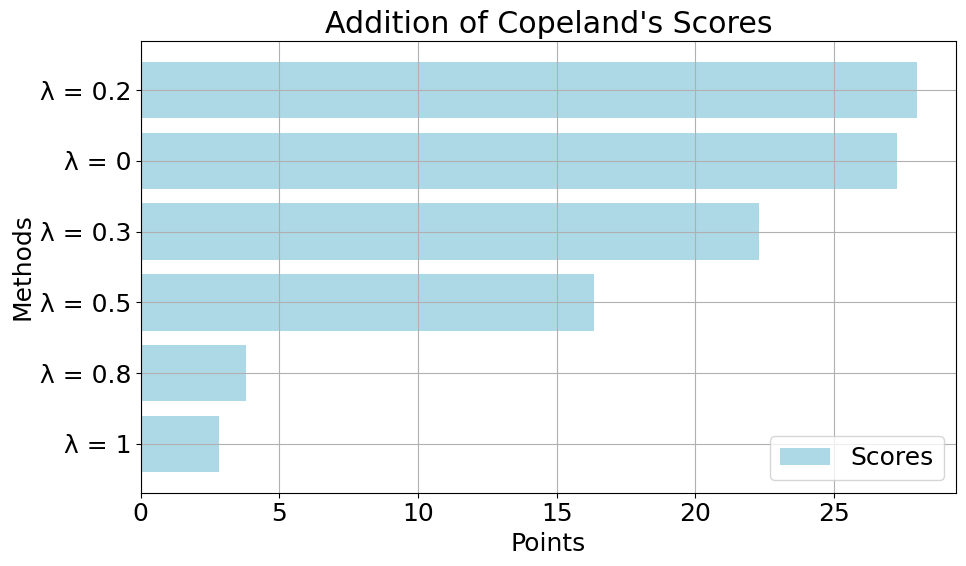}
    \caption{Comparison of Copeland scores for different $\lambda$ values across 100 experiments, in maze dimensions 15x15, 30x30, and 50x50.}
    \label{fig:new_cu_lambda}
    \Description{Copeland Comparison for different $\lambda$ values.}
\end{figure}

In addition, $\lambda$ is an extra parameter introduced in the proposed utility function to balance between the two utility functions. Through 100 experiments conducted across various maze sizes (15x15, 30x30, and 50x50) with 1, 2, 4, 6, 8, and 10 agents, we determined that the optimal value of $\lambda$ in our scenario is 0.2. A more comprehensive comparison of the outcomes across different $\lambda$ values is depicted in Figure \ref{fig:new_cu_lambda}, using Copeland's method (described in Section \ref{sec:result_plots}).

The first utility score of the utility function \ref{eq:utility}, $u_{mnm}$, for the (nearest) frontier cell $f$ is defined as:

\begin{equation}
u_{mnm}(f) = \sum_{i=1}^{n} \text{dist}[(X_{f}, Y_{f}), (X_{r_i}, Y_{r_i})]
\end{equation}

\noindent where, $x_0=(X_{r_i}, Y_{r_i})$ represents the position of agent $i$, $x_n=(X_{f}, Y_{f})$ denotes the position of frontier $f$, and $n$ signifies the number of agents. 


\noindent Also, $u_{jgr}$ for the frontier cell $f$ is defined as:

\begin{equation}
u_{jgr}(f) = \sum_{x_i \in P} \text{Unex}(x_i, r)
\end{equation}

\noindent where $Unex(x, r)$ represents the number of unexplored cells within the agent's view range at cell $x$. $x_i$ denotes the individual elements (cells) of the $path$, designated as the wavefront path, guiding the agent towards the target frontier $f$. 

It is also important to consider that in multi-agent exploration scenarios, there is a possibility that one agent may inadvertently reach the goal intended for another agent. To counter this, our method checks whether the designated goal has been explored in each round. If it has, we repeat the calculation of the next optimal goal for the agent, enhancing overall exploration efficiency.

\subsection{Collision and path blocking avoidance}
\label{subsec:collision_path_block}
The new cost-utility approach also incorporates collision avoidance strategies to navigate around other agents or obstacles. Specifically, if an agent is unable to reach any nearest frontier due to obstacles or blocked paths caused by other agents, according to wavefront, it remains stationary until the obstructing agents move.
Moreover, we implement a mechanism to prevent agents from selecting the same frontiers or goals. This is achieved by excluding each frontier already assigned to an agent. Consequently, each frontier is exclusively designated to a single agent based on its corresponding utility function. It's important to note that we impose this restriction only when the count of unexplored frontiers exceeds or equals the number of agents. The assignment of distinct goals to each agent also serves to mitigate collisions and path-blocking, as the likelihood of agents colliding is reduced.

In Algorithm \ref{alg:new_cost_utility}, we describe the proposed cost-utility for finding the best nearest frontier for each agent, while Algorithm \ref{alg:maze_main_algorithm} describes the main exploration process used in the experiments.

\begin{algorithm}\small
    \caption{Goal Update using Cost-Utility Function}
    \label{alg:new_cost_utility}
    \SetAlgoLined
    \KwIn{agents, total explored map}
    \KwResult{Updates agents' goals}
    Define $frontiers$\;
    
    \ForEach{$a$ in agents}{
        \If{$a$'s goal is unexplored \textbf{and} $frontiers$ $\geq$ agents}{
            Remove $a$'s goal from $frontiers$\;        
        }
    }
    \ForEach{$a$ in agents}{
        \If{$a$'s goal is unexplored}{
            Skip $a$\;
        }
        Calculate nearest frontiers of $a$\ from $frontiers$ using wavefront\;
        
        \If{number of nearest frontiers is 1}{        
            $a$'s goal $\gets$ nearest frontier\;            
        }
        \Else{
            $a$'s goal $\gets$ nearest frontier with maximum utility (\ref{eq:utility})\;
        }
        \If{$frontiers \geq$ agents}{
            Remove $a$'s goal from $frontiers$\;        
        }
    }
\end{algorithm}

\begin{algorithm}\small
    \caption{Main Exploration Process}
    \label{alg:maze_main_algorithm}
    \SetAlgoLined
    \KwResult{Experiment Metrics of Section ~\ref{subsec:experiment_metrics}. }
    Execute Goals Algorithm \ref{alg:new_cost_utility}\;
    Initialize total explored map, rounds, cost exploration metric, round time metric\;
    
    \While{maze coverage $<$ 100\%}{
        rounds $\gets$ rounds + 1\;
        \ForEach{$a$ in agents}{
            $path \gets \text{wavefront}(a, \text{$a$'s goal})$\;
            \If{$path$ is None}{
                $a$ stays in current position\;            
            }
            \Else{
                Move $a$ one cell according to $path$\;
                Update cost exploration metric\;
            }
            Update $a$'s explored \& total explored map\;
        }
        Update Goals using Algorithm \ref{alg:new_cost_utility}\;

        Update round time metric\;
    }
    Calculate other metrics\;
\end{algorithm}

\section{Experimental evaluation}
\label{sec:experimental_eval}

To evaluate the performance of the newly proposed maze exploration method and compare it with other state-of-the-art techniques, we conducted experiments using randomly generated mazes and a varying density of obstacles. We use several metrics to evaluate the performance of the various methods under different conditions. The experiments ran several times and the average scores are reported.

All the code for our experiments and their results is available on a code repository\footnote{\href{https://github.com/manouslinard/multiagent-exploration/}{https://github.com/manouslinard/multiagent-exploration}}.

\subsection{Maze Generation}\label{subsubsec:maze_gen}

For maze generation, we adapted the methodology outlined in \cite{mazegen}. This approach involves moving an agent randomly in orthogonal directions (up, down, left, and right) within a grid to construct the maze. Additionally, we extended this approach by introducing a probability factor for each obstacle cell. This probability determines whether an obstacle cell should remain or be converted into empty space, allowing the creation of mazes with varying levels of complexity. The function also allows for the specification of maze dimensions.

The maze is represented as a 2D array with three cell states: free space (0), obstacles (1), and agent cells (2). In the explored map, cells outside agents' view ranges are marked as -1 (unexplored). When an agent views its surroundings, cells within its range are marked using 0 for free space and 1 for obstacles. During the experiments, all agent positions (cells with value 2) are known, ensuring visibility of agent locations within the maze.


For our experiments, we generated mazes with dimensions of 15x15, varying the obstacle probabilities to create more and less complex environments, set at 85\% and 15\%, respectively. This range of maze configurations enabled us to comprehensively evaluate the algorithms' performance under various conditions. The choice of a 15x15 maze size is particularly noteworthy as it represents an intermediate dimension in maze exploration research. Recent studies, such as the HEDAC study \cite{hedac_maze}, utilized maze dimensions of 10x10 and 20x20 for comparative analysis with other methods. So, opting for a 15x15 maze aligns with current trends, providing an optimal balance for our experimental objectives.

\begin{figure}[h]
    \centering
    \includegraphics[width=0.45\textwidth]{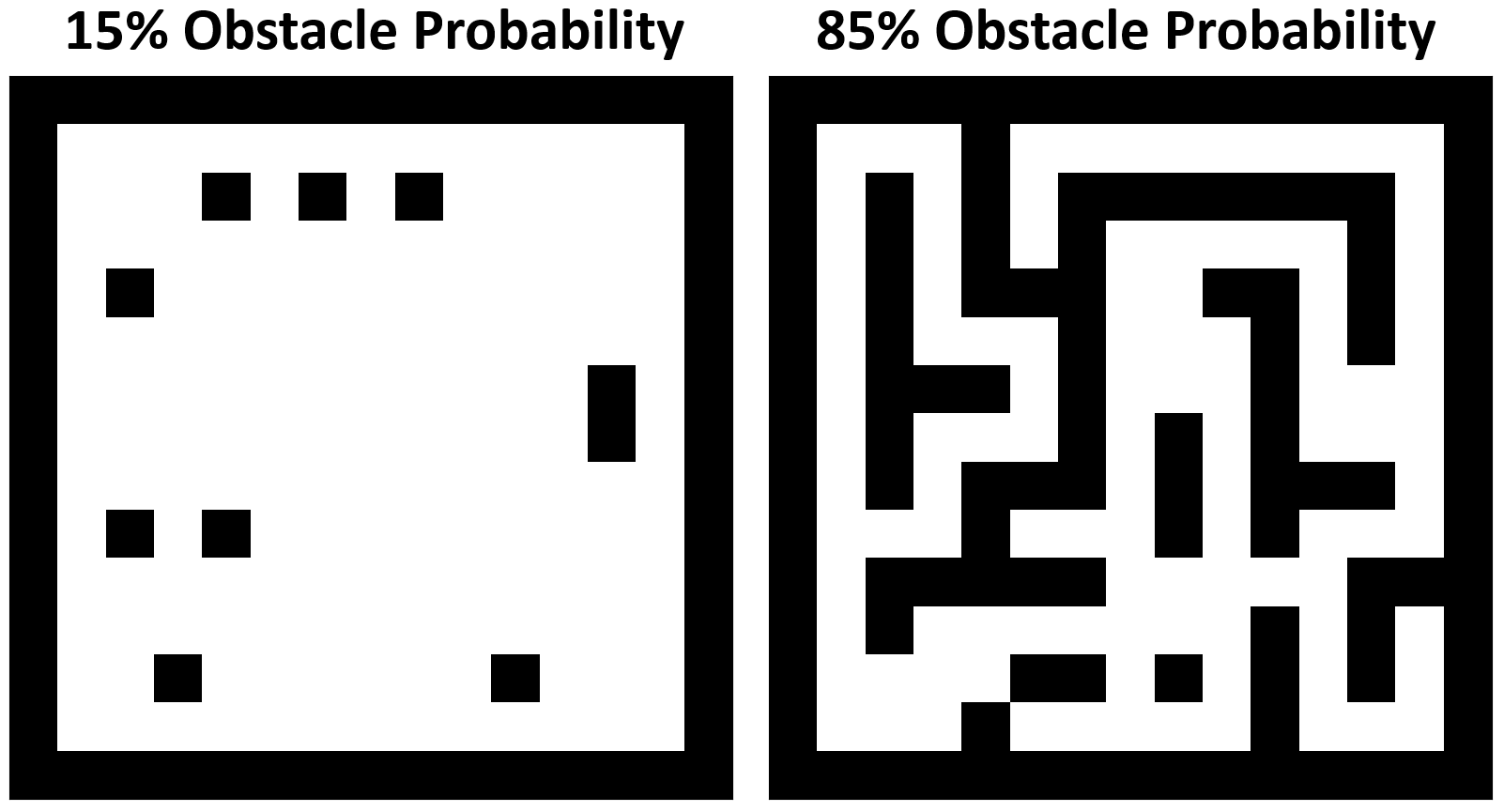}
    \caption{Random 15x15 mazes from the experiments.}
    \label{fig:mazes}
    \Description{Randomly generated 15x15 mazes from the experiments.}
\end{figure}

To ensure robustness in our evaluation, we generated new mazes for each experiment, providing a fresh perspective on the algorithm's efficiency. Figure \ref{fig:mazes} illustrates two randomly generated mazes, one with an obstacle probability of 15\% and the other with 85\%, showcasing the diversity in maze complexity explored in our experiments.

\subsection{Baseline methods}\label{subsec:baseline_methods}

The baseline methods used to compare the new algorithm are described in Table \ref{tab:baseline_table}. Each method has been implemented from scratch based on the corresponding papers.

\begin{table}
    \centering
    \begin{tabular}{|p{0.2\linewidth}|p{0.7\linewidth}|}
    \hline
    \multicolumn{1}{|c|}{\textbf{Name}} & \multicolumn{1}{c|}{\textbf{Description}} \\
    \hline
    HEDAC & Utilizes potential fields to explore the maze \cite{hedac_maze}. \\
    \hline
    Nearest-Front & Selects the closest frontier to the agent \cite{yamauchi1998frontier}. \\
    \hline
    CU-BSO & Uses wavefront propagation \& selects the frontier with less robots near it \cite{cu_bso}. \\
    \hline
    CU-MNM & Cost-utility method considering the distances of frontiers from all agents \cite{cu_mnm_paper}. \\
    \hline
    CU-JGR & Cost-utility method estimating the expected explored cells in the target cell \cite{cu_jgr}. \\
    \hline
    Flood-Fill & A simple flood fill approach for maze exploration. \\
    \hline
    \end{tabular}
    \caption{Baseline methods implemented \& used for comparison with new cost-utility method.}
    \label{tab:baseline_table}
\end{table}

Some methods also require specific parameter settings for optimal performance. For instance, in HEDAC, there is an "iterations" parameter, as described in the original paper \cite{hedac_maze}, which determines the number of times the attractive force should be recalculated to produce the final attractive force for each cell. Through experimentation, we determined that setting this parameter to 100 produced the best outcomes. Lower values resulted in agents becoming trapped in certain regions without finding a target, hindering the exploration progress, while higher values substantially prolonged computation time. Additionally, HEDAC defines a parameter denoted as $a$, which performs optimally when set to 10. Also, the anti-collision (AC) condition outlined in the original HEDAC paper is set to ON. This means that if an agent has information that one of the other agents is currently standing on one of its neighboring nodes, the agent does not consider this node for its next position. Furthermore, in our implementation, we extend the original HEDAC approach by setting the agent's view range to two blocks.

\begin{figure}
  \centering
  \includegraphics[width=0.47\textwidth]{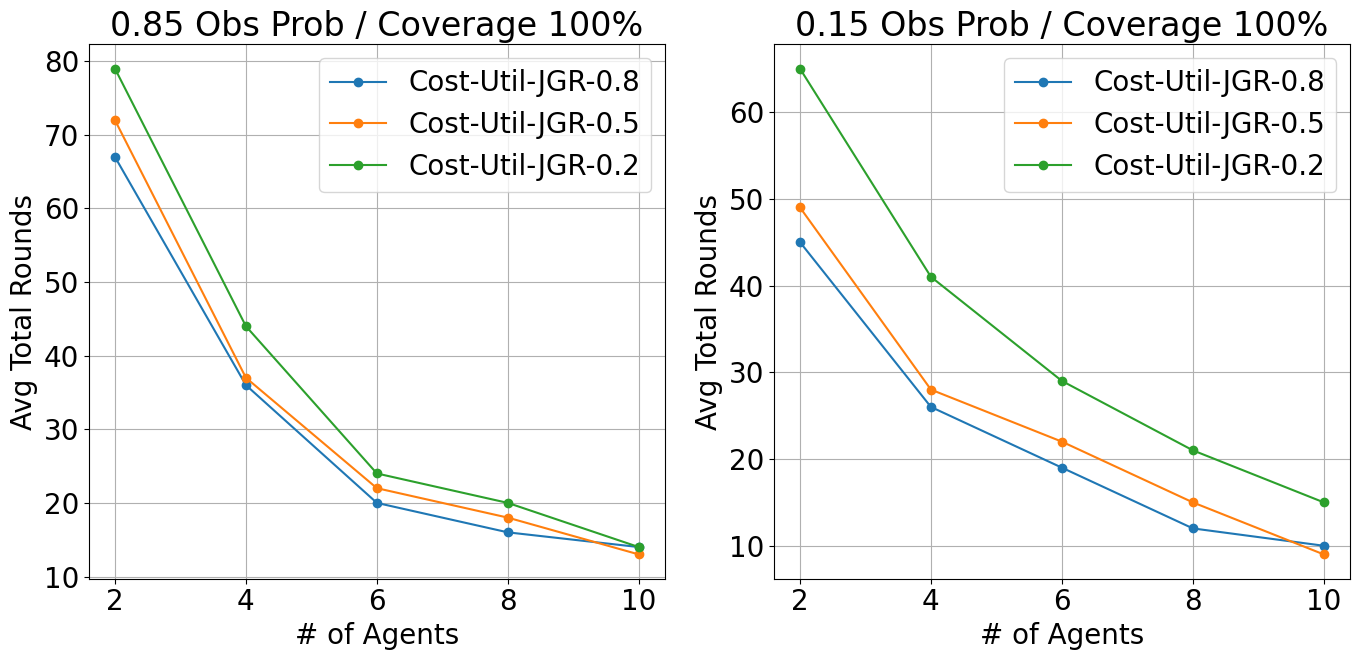} 
  \caption{Comparison of Total (Exploration) Rounds (\ref{subsubsec:exploration_rounds}) for different $\lambda_{jgr}$ values of CU-JGR in 15x15 mazes.}
  \label{fig:lambda_comp}
  \Description{Comparison of Total (Exploration) Rounds of $\lambda_{jgr}$.}
\end{figure}

The CU-JGR algorithm \cite{cu_jgr} also requires the setting of a parameter $\lambda_{jgr}$. Through experimentation, we determined that setting $\lambda_{jgr}$ to 0.8 resulted in more efficient maze exploration, reducing the average number of exploration rounds. Figure \ref{fig:lambda_comp} illustrates the results obtained for different $\lambda_{jgr}$ values.

The baseline Flood Fill algorithm draws inspiration from the principles outlined in \cite{tjiharjadi2022design}, utilizing flood fill to determine the distances of the agent from all cells within the maze. This algorithm involves executing flood fill ascending from each frontier and reaching the agents. Subsequently, agents select the frontier with the smallest distance calculated via flood fill. The implementation of this selection process is close to the approach used in CU-BSO \cite{cu_bso}, which employs a wavefront-based method for a similar purpose.

\subsection{Evaluation metrics}\label{subsec:experiment_metrics}
For method comparison, we utilized evaluation metrics proposed by Yan et al. \cite{yan2015metrics}. Additionally, we tracked the number of exploration rounds to measure the times that the exploration process is repeated. More specifically, the evaluation metrics employed in this study include:
i) Exploration Rounds, ii) Exploration Time, iii) Exploration Cost, iv) Exploration Efficiency, and v) Map Quality. 

\subsubsection{Exploration Rounds}\label{subsubsec:exploration_rounds}
Exploration rounds ($R$) represent the repetitive cycles of the exploration process, encapsulating both information exchange among agents and the algorithm's repetition until maze completion. These iterations serve as a measure of the times the agents have communicated with each other, as information sharing occurs in each round. Exploration iterations conclude when the entire maze has been fully explored.

\subsubsection{Exploration Cost}
\label{subsubsec:exploration_cost}
The exploration cost is based upon user specifications, encompassing factors such as energy consumption by computational resources (e.g., CPU, RAM, and network bandwidth), as well as robot-related expenses such as acquisition, handling, and maintenance costs. Their definition of the exploration cost metric requires summing the distances traveled by each robot within the fleet. This metric is calculated using the following formula:

\begin{equation}
explorationCost(n) = \sum_{i=1}^{n} d_i
\label{eq:exp_cost}
\end{equation}

\noindent where $n$ is the number of robots in the fleet, and $d_i$ is the distance traveled by robot $i$.

\subsubsection{Exploration Efficiency}
\label{subsubsec:exploration_efficiency}
Exploration efficiency correlates with the quantity of environmental information acquired and inversely relates to the costs incurred by the robot fleet.
So it can be described with the following formula:

\begin{equation}
explorationEfficiency(n) = \frac{M}{explorationCost(n)}
\label{eq:exp_eff}
\end{equation}
\noindent where $n$ is the number of robots in the fleet and $M$ is the total explored cells.

\subsubsection{Exploration Time}
\label{subsubsec:exploration_time}
The exploration time metric measures the total time required to complete an exploration mission for a robot fleet. The timer initiates when at least one robot within the fleet commences exploration and halts when the fleet collectively achieves a predetermined percentage of terrain exploration, in our case 100\%. This time metric is quantified using wall-clock time, reflecting the duration in days, hours, minutes, and seconds that the fleet has dedicated to the exploration endeavor.

In our experiments, the exploration time is calculated by finding the average time of each round and multiplying it by the number of rounds. Since the agents operate sequentially (one agent after the other performs a step in each round) we calculate the average round time by dividing the time needed for all agents to make a step to the number of agents.

\begin{equation}
explorationTime(n) = \frac{\sum_{i=1}^{n} t_{s_i}}{n} \cdot R
\label{eq:exp_time}
\end{equation}
\noindent where $t_{s_i}$ is the time needed for an agent to make a step.

\subsubsection{Map Quality}
\label{subsubsec:map_quality}
The map quality is defined as the overlap of the explored map and the ground truth map as a ratio to the total area of the ground truth map P:
\begin{equation}
mapQuality = \frac{M - A(\text{mapError})}{P}
\label{eq:map_quality}
\end{equation}

\noindent where $M$ are the total explored cells, $P$ the total area of the ground truth map, and $A(\text{mapError})$ is the area occupied by the error cells. The error cells are the cells in the explored map that have a different value from the corresponding cell in the ground truth map. Overall in the experiments, the $mapQuality$ metric is always at 100\% due to the efficiency of the algorithms and the design of the mazes, which facilitate agents' access to all areas for exploration.

\subsection{Hardware description}
\label{subsec:hardware_desc}
The final experiments were conducted on a PC running Windows 11 equipped with an Intel(R) Core(TM) i5-6600K CPU. To expedite the experiment process, we utilized the multiprocessing library in Python. 
The results of the experiments are presented in Section \ref{sec:result_plots}.

\section{Results and Discussion}\label{sec:result_plots}

The results comprise 1000 experiments conducted with varying numbers of agents, specifically 1, 2, 4, 6, 8, and 10. Each agent configuration explored mazes of dimensions 15x15 with obstacle probabilities set at 85\% and 15\%. These experiments encompassed the utilization of both baseline methods and the newly proposed cost-utility approach, referred to as `New-CU-DIFFGOAL-PATH' in the resulting plots. The presented plots showcase the average scores and deviations derived from these 1000 experiments for each agent group and exploration method.

Despite the large number of experiment repetitions that significantly reduce the deviation around average scores, some of the methods as shown in Figures \ref{fig:avg_rounds}, \ref{fig:avg_cost}, \ref{fig:avg_eff} and \ref{fig:avg_time} perform comparably. 
Given the occasional proximity of results between methods, Copeland's method \cite{copeland3, copeland1} was employed to ascertain the comparative effectiveness of each approach. Copeland's method is based on pairwise comparisons of the average scores obtained from each method, with the best (method) obtaining more points/votes. This voting system determines which method gained more votes, thereby establishing its superiority across a wide range of experiments (e.g. with a varying number of agents or with various obstacle densities).

\subsection{Results of Exploration Rounds}
\label{subsec:res_expl_rounds}
\begin{figure}
  \centering
  \includegraphics[width=0.47\textwidth]{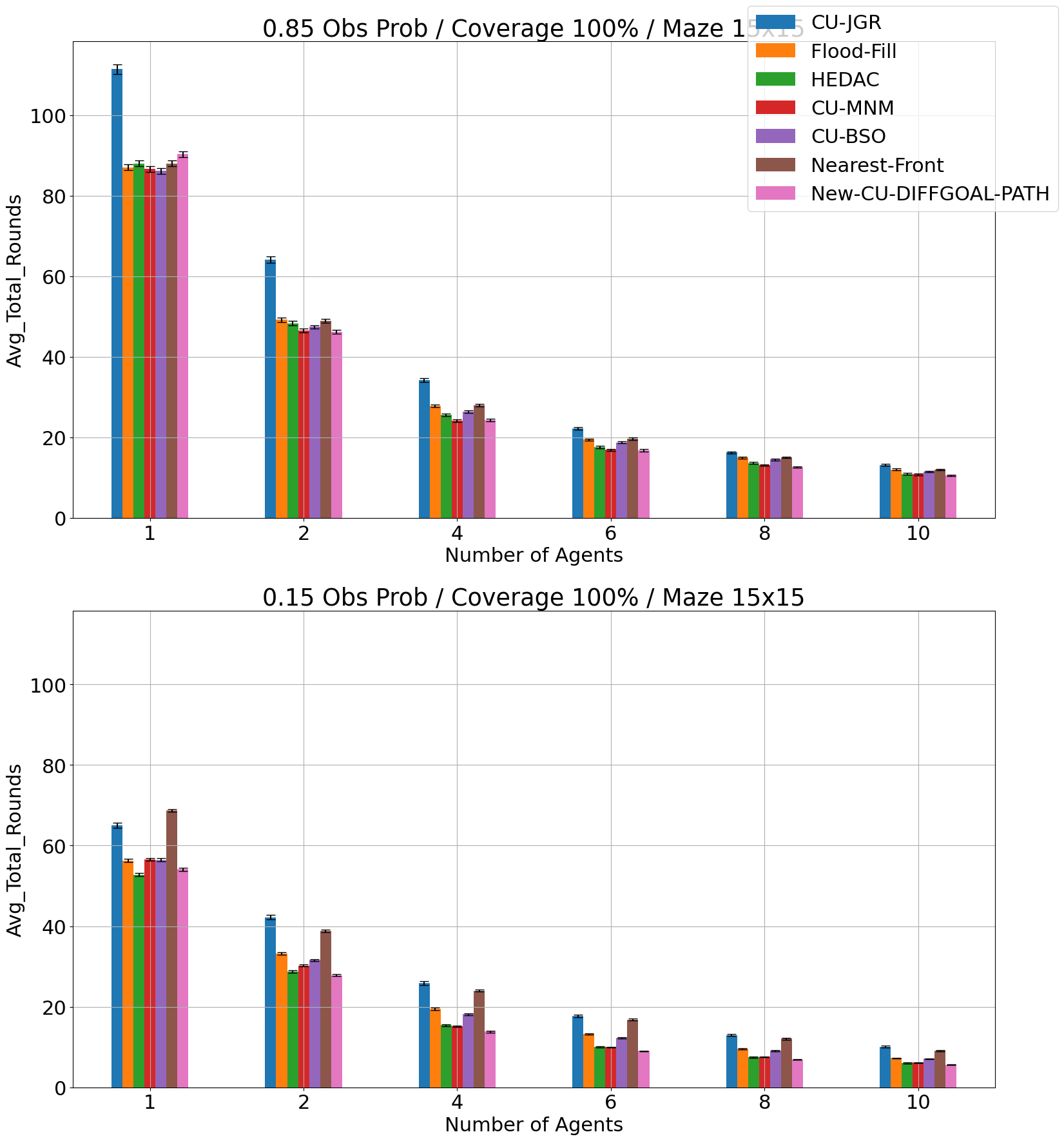}
  \caption{Average Total (Exploration) Rounds throughout 1000 experiments.}
  \label{fig:avg_rounds}
  \Description{Average Total (Exploration) Rounds throughout 1000 experiments.}
\end{figure}

\begin{figure}
  \centering
  \includegraphics[width=0.47\textwidth]{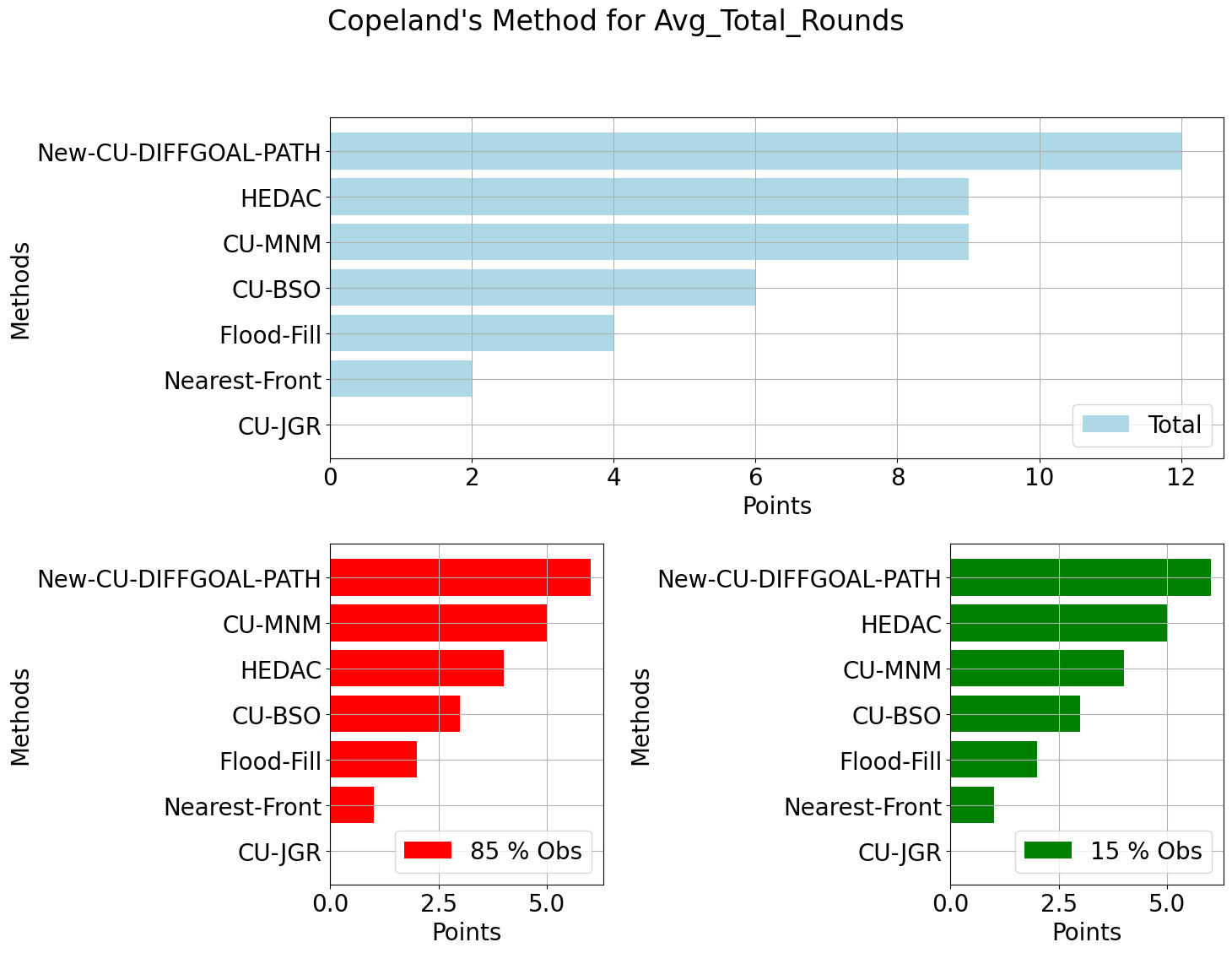}
  \caption{Copeland's Voting Method results for Average Total (Exploration) Rounds.
  }
  \Description{Copeland's Voting Method results for Average Total (Exploration) Rounds.}
  \label{fig:avg_cop_rounds}
\end{figure}

Figure \ref{fig:avg_rounds} presents the average exploration rounds obtained from 1000 experiments conducted on mazes with dimensions of 15x15 and obstacle probabilities set at 85\% and 15\%. The results demonstrate the efficiency of the new cost-utility approach in exploration rounds, consistently performing as well as or better than other compared methods. To further analyse performance as the number of agents increases, Copeland's method is employed and depicted in Figure \ref{fig:avg_cop_rounds}.

In the comparison plot (Fig. \ref{fig:avg_cop_rounds}), the new cost-utility method received the highest number of votes across all maze configurations. This emphasizes the impressive performance of the new method, particularly in terms of exploration rounds, which translates to reduced communication between agents.

\subsection{Results of Exploration Cost}
\label{subsec:res_expl_cost}
\begin{figure}
  \centering
  \includegraphics[width=0.47\textwidth]{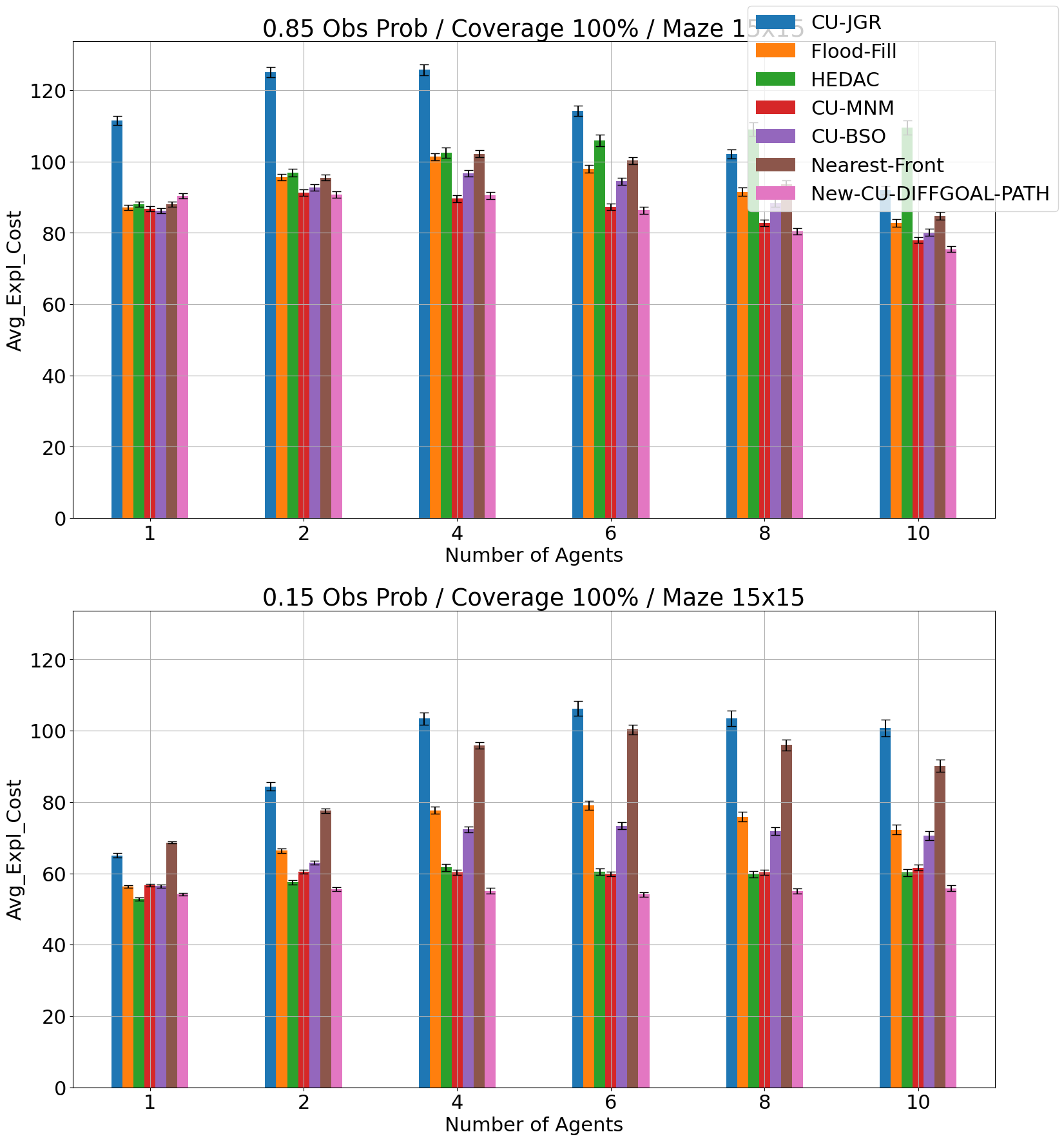}
  \caption{Average Exploration Cost throughout 1000 experiments.}
  \Description{Average Exploration Cost throughout 1000 experiments.}
  \label{fig:avg_cost}
\end{figure}

Figure \ref{fig:avg_cost} illustrates the average exploration cost metric when different numbers of agents are used with different methods. Once more, the new cost-utility method demonstrates competitive performance compared to the baseline methods.

\begin{figure}
  \centering
  \includegraphics[width=0.47\textwidth]{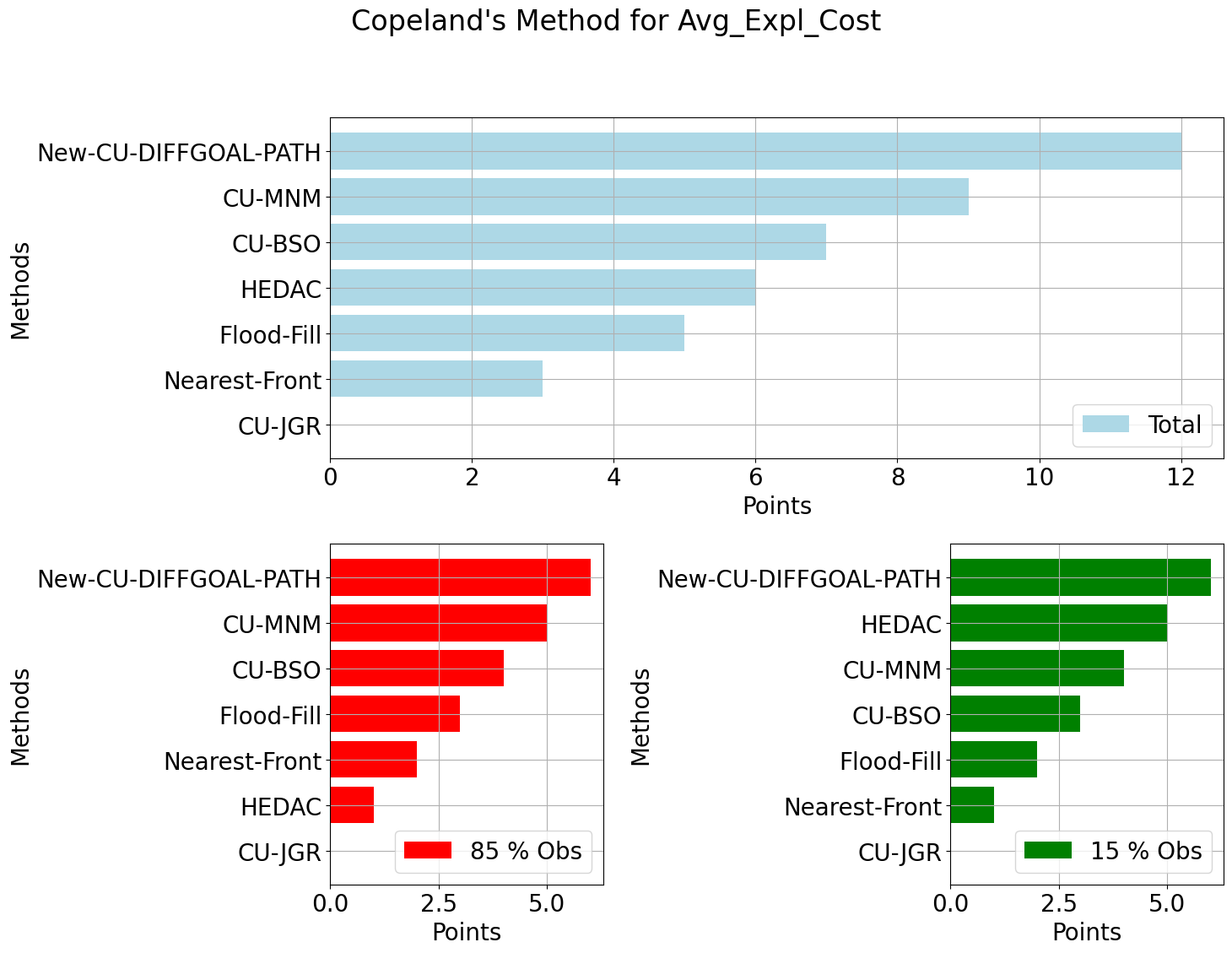}
  \caption{Copeland's Voting Method results for Average Exploration Cost.
  }
  \Description{Copeland's Voting Method results for Average Exploration Cost.}
  \label{fig:avg_cop_cost}
\end{figure}

This observation is further supported by the results depicted in Figure \ref{fig:avg_cop_cost}. The new cost-utility method consistently achieves the highest scores in all maze variations. This result underscores once again the efficiency of the frontier selection process, resulting in reduced duplication of efforts by agents and minimized traversal distances, as indicated by the exploration cost metric.

\subsection{Results of Exploration Efficiency}
\label{subsec:res_expl_eff}
\begin{figure}
  \centering
  \includegraphics[width=0.45\textwidth]{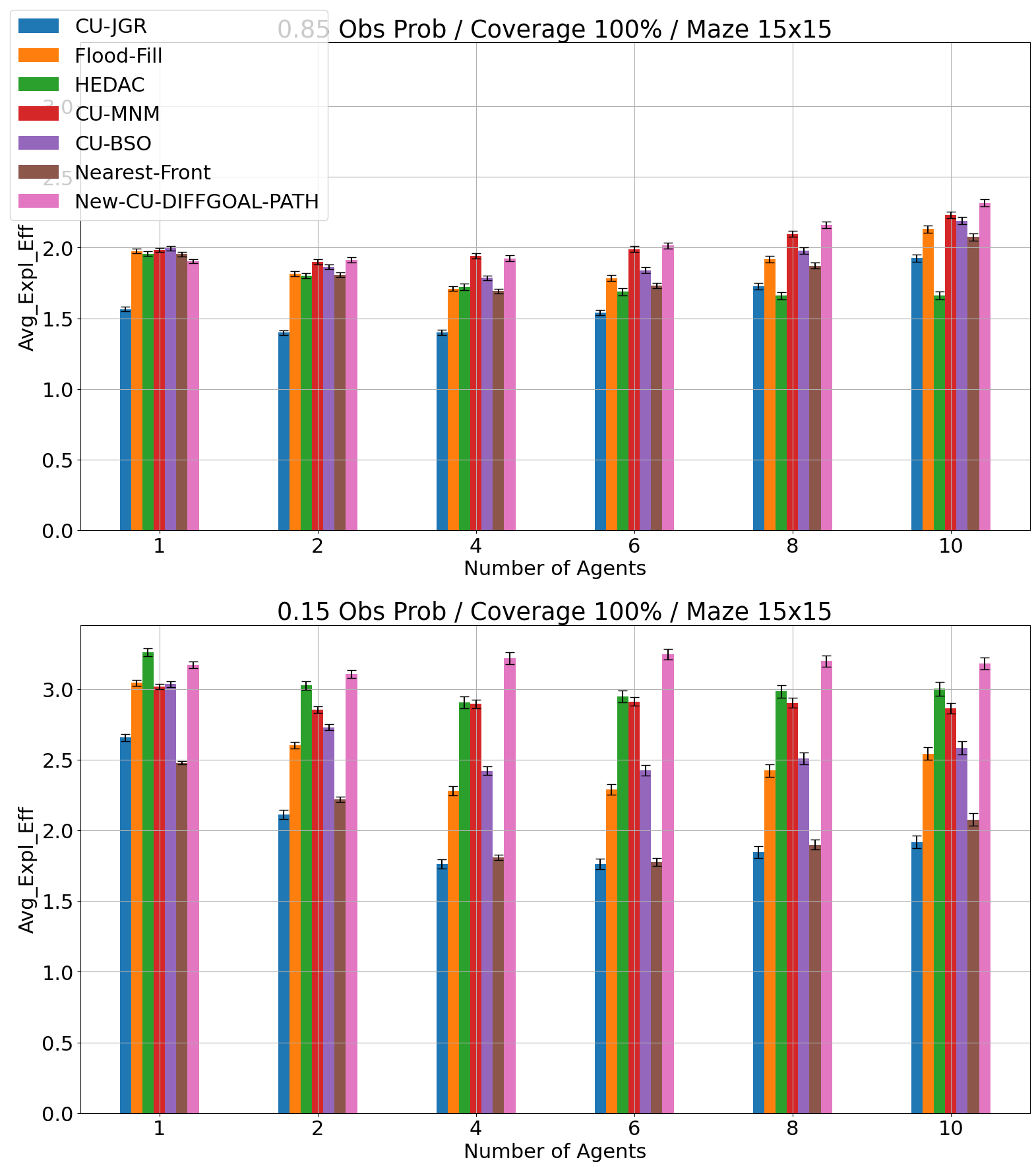}
  \caption{Average Exploration Efficiency throughout 1000 experiments.}
  \label{fig:avg_eff}
  \Description{Average Exploration Efficiency throughout 1000 experiments.}
\end{figure}

Figure \ref{fig:avg_eff} illustrates the average exploration efficiency metric for varying numbers of agents across all experiments and methods. Once again, the new cost-utility method demonstrates competitiveness against the baseline methods, often surpassing them in efficiency.

\begin{figure}
  \centering
  \includegraphics[width=0.47\textwidth]{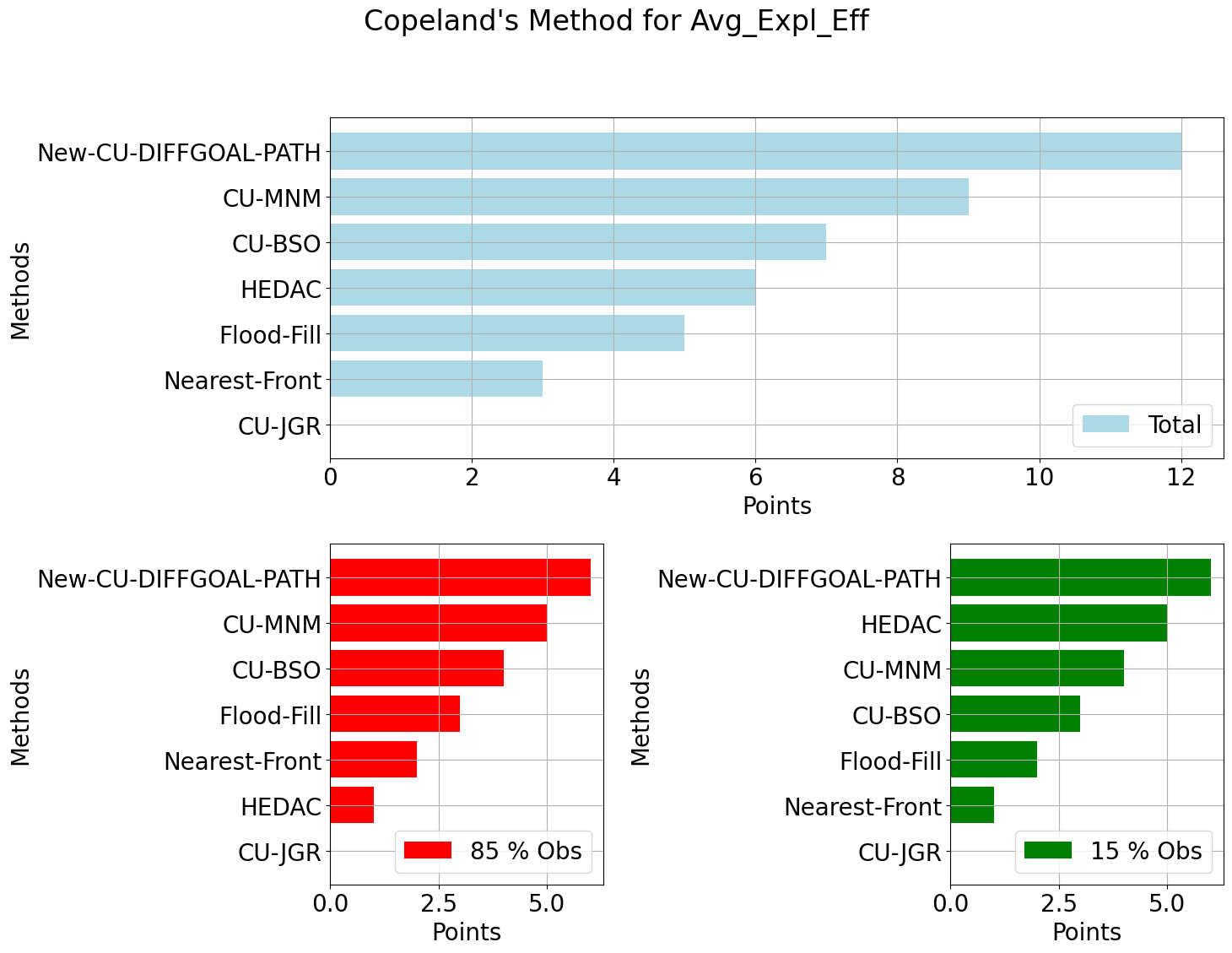}
  \caption{Copeland's Voting Method results for Average Exploration Efficiency.
  }
  \Description{Copeland's Voting Method results for Average Exploration Efficiency.}
  \label{fig:avg_cop_eff}
\end{figure}

This trend is further reinforced by the results obtained from Copeland's Method, as depicted in Figure \ref{fig:avg_cop_eff}. Once again, the new cost-utility method consistently secures the top position in all examined mazes.

Furthermore, it's noteworthy that these results closely mirror those obtained in the exploration cost metric, given that exploration efficiency formula (\ref{eq:exp_eff}) is essentially the reverse application of the exploration cost formula (\ref{eq:exp_cost}).

\subsection{Results of Exploration Time}
\label{subsec:res_exp_time}
\begin{figure}
  \centering
  \includegraphics[width=0.47\textwidth]{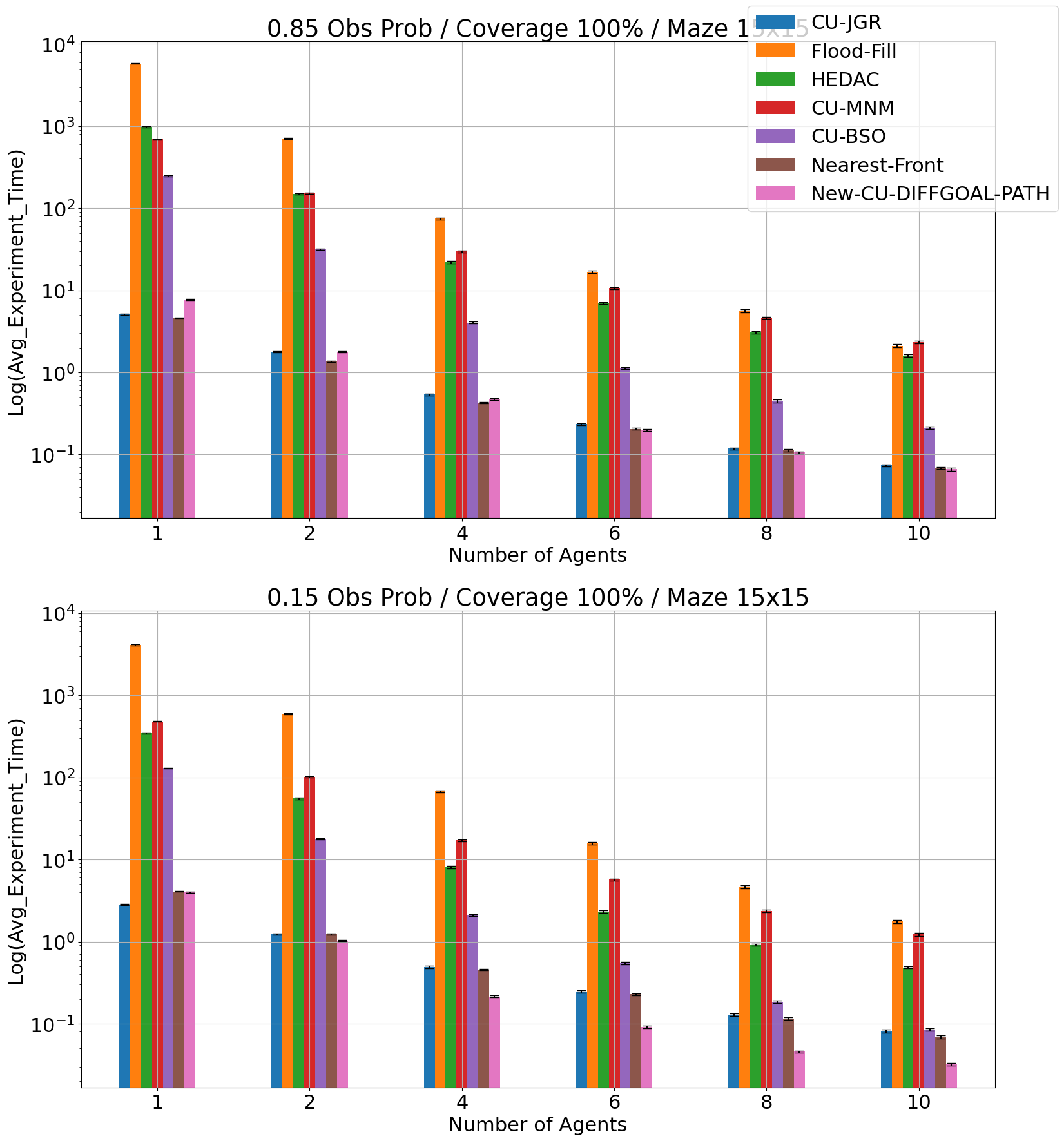}
  \caption{Average Experiment (Exploration) Time throughout 1000 experiments in logarithmic scale.}
  \Description{Average Experiment (Exploration) Time throughout 1000 experiments in logarithmic scale.}
  \label{fig:avg_time}
\end{figure}

Figure \ref{fig:avg_time} depicts the average exploration time for varying numbers of agents across all experiments and methods in logarithmic scale, for better visualization. The new method demonstrates very competitive performance to all other methods across all agent numbers. As the number of agents increases, the exploration time tends to stabilize across all methods due to the relatively small maze size. However, some baseline methods, such as flood fill, still exhibit inferior performance compared to others, underscoring the efficiency of the new cost-utility approach.

\begin{figure}
  \centering
  \includegraphics[width=0.47\textwidth]{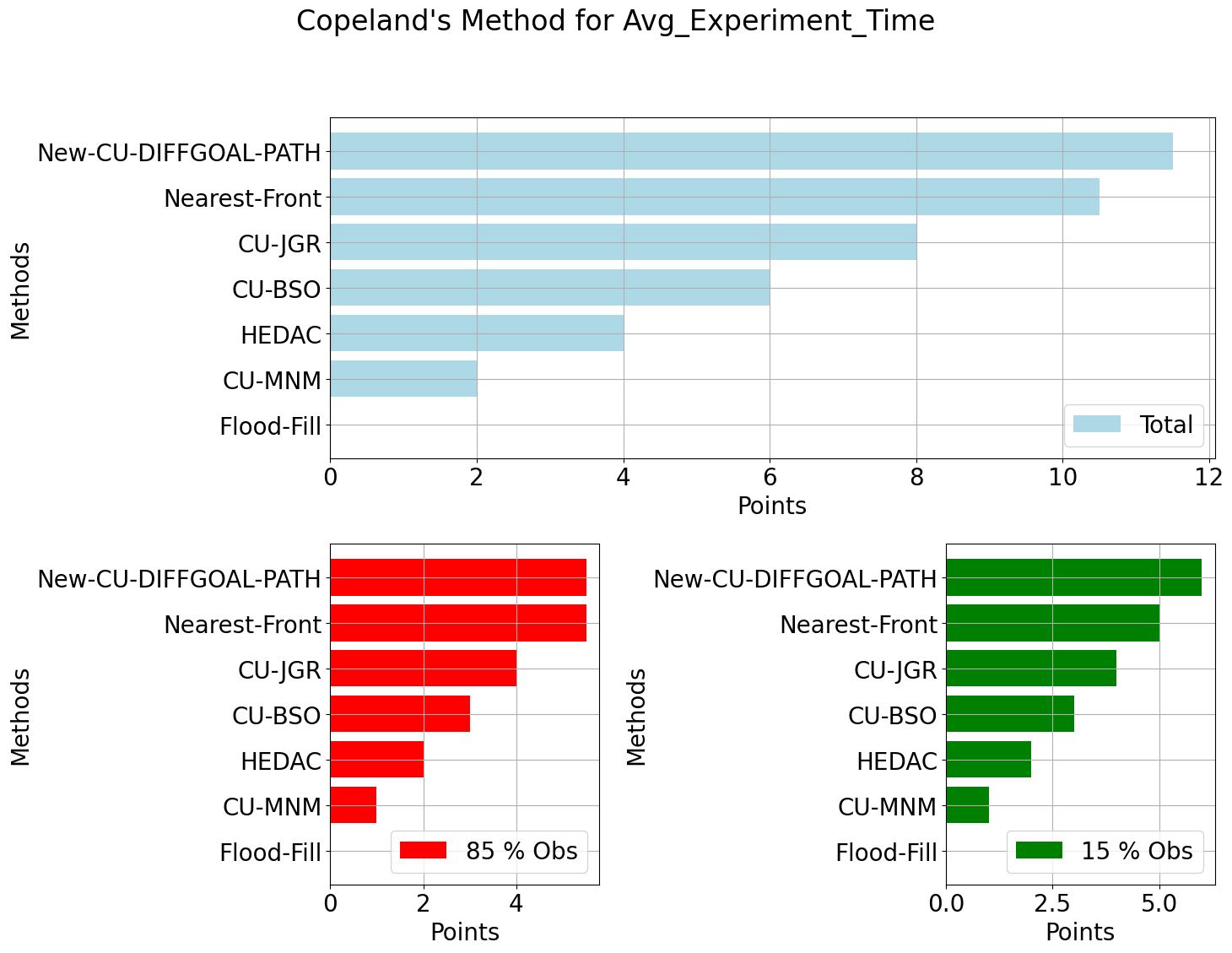}
  \caption{Copeland's Voting Method results for Average Experiment (Exploration) Time.
  }
  \Description{Copeland's Voting Method results for Average Experiment (Exploration) Time.}
  \label{fig:avg_cop_time}
\end{figure}

The differences in exploration time are highlighted in Figure \ref{fig:avg_cop_time}. The proposed algorithm exhibits the best exploration time, especially in mazes with fewer obstacles. In more complex mazes, it performs comparably to the nearest frontier algorithm. The additional time needed to compute the two cost functions is balanced by the faster execution of wavefront for finding the nearest frontiers and their distance from the agent. Also, the new approach avoids the limitations that stem from the simplicity of the nearest frontier algorithm in finding optimal frontier choices and revisiting cells.


Additionally, we need to consider that in real-life scenarios, the computation time of the algorithm may be significantly shorter than the movement time, especially for vehicular robots navigating through areas or buildings. Therefore, prioritizing more computation to minimize movement distance (or in our metrics, the exploration cost) can be more beneficial for overall exploration efficiency.

\section{Conclusions and Next Steps}
\label{sec:conclusions}

The performance of the new maze exploration method outperformed the state-of-the-art methods used for comparison in most of the metrics. It ranked first in the number of exploration rounds, the total exploration cost, and the efficiency of exploring the whole maze with minimum exploration cost. In terms of computational complexity (i.e. exploration time), the new maze exploration method consistently ranked first across all examined mazes. It shared the top position with the nearest frontier method only in the most complex mazes, which is very impressive considering the nearest frontier is computationally simple. In addition, it avoids selecting the same targets for multiple agents.

Our next research steps involve integrating successful elements from other methods, such as HEDAC into the utility function to further improve performance. Conducting experiments in larger mazes with dynamic sizes and a broader range of complexities will allow for better configuration of the utility function's $\lambda$ parameter and a deeper understanding of the algorithm's performance in more expanded settings. We also plan to incorporate broadcast range limitations to simulate realistic robot interactions and explore decentralized approaches that assume partial knowledge of the explored map for each agent. Moreover, applying the methods from this work in real-world environments holds potential for delivering valuable insights into their performance and efficiency when faced with practical challenges. This will also require taking into account real-world considerations, such as broadcast range and communication costs, which will drive further innovations and adjustments, thereby expanding their applicability. Finally, developing new metrics that quantify agents' energy consumption could provide deeper insights into algorithmic efficiency.

\section*{Acknowledgment}
The research leading to these results has received funding from the European Union’s Horizon Europe research and innovation programme under grant agreement No 101073876 (Ceasefire). This publication reflects only the authors views. The European Union is not liable for any use that may be made of the information contained therein.

\bibliographystyle{ACM-Reference-Format}
\bibliography{sample-base}

\end{document}